\documentclass[a4paper,11pt]{article}
\usepackage{pos}

\title{Quark orbital angular momentum in the proton
from a twist-3 generalized parton distribution}
\ShortTitle{Quark orbital angular momentum in the proton
from a twist-3 GPD}

\author*[a]{M.~Engelhardt}
\author[b]{N.~Hasan}
\author[c,d]{S.~Krieg}
\author[e]{S.~Liuti}
\author[f]{S.~Meinel}
\author[g]{J.~Negele}
\author[g]{A.~Pochinsky}
\author[c,d]{M.~Rodekamp}
\author[h]{S.~Syritsyn}

\affiliation[a]{Department of Physics, New Mexico State University,
Las Cruces, NM 88003, USA}
\affiliation[b]{Bergische Universit\"at Wuppertal, 42119 Wuppertal, Germany}
\affiliation[c]{J\"ulich Supercomputing Centre \&\, Institute for Advanced
Simulation \&\, Center for Advanced Simulation and Analytics (CASA),
Forschungszentrum J\"ulich, 54245 J\"ulich, Germany}
\affiliation[d]{Helmholtz-Institut f\"ur Strahlen- und Kernphysik,
Rheinische Friedrich-Wilhelms-Universit\"at Bonn, \linebreak
53115 Bonn, Germany}
\affiliation[e]{Department of Physics, University of Virginia,
Charlottesville, VA 22904, USA}
\affiliation[f]{Department of Physics, University of Arizona, Tucson,
AZ 85721, USA}
\affiliation[g]{Center for Theoretical Physics, Massachusetts Institute of
Technology, Cambridge, MA 02139, 
USA}
\affiliation[h]{Department of Physics and Astronomy, Stony Brook University,
Stony Brook, NY 11794, USA}

\emailAdd{engel@nmsu.edu}

\abstract{\vspace{0.3cm}
Quark orbital angular momentum in the proton is evaluated via
a Lattice QCD calculation of the second Mellin moment of the twist-3
generalized parton distribution $\widetilde{E}_{2T} $ in the forward
limit. The connection between this approach to quark orbital angular
momentum and approaches previously utilized in Lattice QCD calculations,
via generalized transverse momentum-dependent parton distributions and
via Ji's sum rule, is reviewed. This connection can be given in terms
of Lorentz invariance and equation of motion relations. The calculation
of the second Mellin moment of $\widetilde{E}_{2T} $ proceeds via a
finite-momentum proton matrix element of a quark bilocal operator with
a straight-line gauge connection and separation in both the longitudinal
and transverse directions. The dependence on the former component serves
to extract the second Mellin moment, whereas the dependence on the latter
component provides a transverse momentum cutoff for the matrix element.
Furthermore, a derivative of the matrix element with respect to momentum
transfer in the forward limit is required, which is obtained using a
direct derivative method. The calculation utilizes a clover fermion
ensemble at pion mass $317\, \mbox{MeV} $. The resulting quark orbital
angular momentum is consistent with previous evaluations through
alternative approaches, albeit with greater statistical uncertainty
using a comparable number of samples. \vspace{0.5cm}}

\FullConference{25th International Spin Physics Symposium (SPIN 2023)\\
 24-29 September 2023\\
 Durham, NC, USA\\}

\begin{document}
\maketitle

\section{Introduction}
The orbital angular momentum carried by quarks inside the proton
constitutes one of the pieces of the proton spin puzzle -- the
question of how the spin of the proton is composed of the spins
and orbital angular momenta of its quark and gluon constituents.
This question has prompted enduring efforts in hadronic physics,
sparked by the initial EMC experiments \cite{emc1,emc2} that revealed that
one cannot simply explain proton spin in terms of valence quark spin.

The objective of quantifying quark orbital angular momentum already
meets challenges at the conceptual level as a consequence of gauge
invariance, which prevents one from unambiguously separating the
quark and gluon degrees of freedom. Quark fields are intrinsically
linked to gluon fields, and consequently any construction of quark orbital
angular momentum includes gluonic effects. The partition of orbital
angular momentum in the proton into a quark and a gluon contribution
is therefore a matter of definition. Whereas, in principle, a continuum
of definitions is possible, two stand out prominently in discussions of
the proton spin puzzle, namely, the Ji definition \cite{jidecomp} and
the Jaffe-Manohar definition \cite{jmdecomp,hatta}.

The quark-gluon structure of hadrons is encoded in parton distribution
functions -- generalized parton distributions (GPDs), revealing transverse
position structure along with longitudinal momentum structure; transverse
momentum-dependent parton distributions (TMDs), revealing transverse
momentum structure along with longitudinal momentum structure; and,
overarching the aforementioned, generalized TMDs (GTMDs), which contain
GPDs and TMDs as limits. In particular, GTMDs furnish mixed transverse
position and momentum information, and are therefore suited for a direct
partonic definition of longitudinal orbital angular momentum \cite{lorce}.

Altogether, three avenues to evaluate longitudinal quark orbital angular
momentum in a longitudinally polarized proton from its parton distribution
functions have been constructed:
\begin{itemize}
\item {\bf Ji's sum rule:} The total longitudinal quark angular momentum
can be derived from the quark GPD combination $H+E$ \cite{jidecomp}, whereas
the longitudinal quark spin is given by the quark GPD $\widetilde{H} $.
One can thus obtain the longitudinal orbital angular momentum indirectly,
by taking the difference (denoting the longitudinal direction as the
3-direction),
\begin{equation}
L_3 = J_3 -S_3 = \frac{1}{2} \int dx\, x (H+E)
-\frac{1}{2} \int dx\, \widetilde{H} \ ,
\label{jisr}
\end{equation}
where $x$ denotes the quark momentum fraction and all GPDs are evaluated
in the forward limit. This relation yields specifically the orbital angular
momentum according to the Ji definition.
\item {\bf Twist-2 GTMD $F_{14} $}: As already mentioned above, direct
access to longitudinal quark orbital angular momentum can be obtained
through a GTMD \cite{lorce}, named $F_{14} $ in the nomenclature of
\cite{mms},
\begin{equation}
L_3 = -\int dx \int d^2 k_T \ \frac{k_T^2}{M^2 } \, F_{14} \ \ ,
\label{lf14}
\end{equation}
where $k_T $ denotes the quark transverse momentum and the GTMD is again
evaluated in the forward limit. This approach allows one to evaluate
quark orbital angular momentum according to both the Ji and the
Jaffe-Manohar definitions (and, more generally, a continuous interpolation
between the two), since the QCD matrix element definition of $F_{14} $ is
based on a bilocal quark operator including a transverse separation, which
allows for a choice of the shape of the gauge connection between the
quark operators. A staple-shaped gauge connection path, as used in the
standard definition of TMDs, yields Jaffe-Manohar quark orbital angular
momentum \cite{hatta,burk}; a straight-line path yields Ji quark orbital
angular momentum \cite{jist,burk,eomlir}.
\item {\bf Twist-3 GPD $\widetilde{E}_{2T} $:} A third way to access
quark orbital angular momentum specifically according to the Ji definition
is through a twist-3 GPD, named $\widetilde{E}_{2T} $ in the nomenclature of
\cite{mms}. $\widetilde{E}_{2T} $ is related to the combination of
orbital angular momentum and spin $L_3 +2S_3 $ \cite{lpproc,eomlir,eomlong},
and one can thus obtain $L_3 $ by subtracting twice the spin $S_3 $,
\begin{equation}
L_3 = (L_3 +2S_3 ) - 2S_3 = 
-\int dx\, x \widetilde{E}_{2T} - \int dx\, \widetilde{H} \ ,
\label{e2ttildeapp}
\end{equation}
where again all GPDs are evaluated in the forward limit. It should
be noted that a version of this approach was first advanced by
M.~Polyakov and collaborators \cite{polyakov1,polyakov2}, denoting
the relevant twist-3 GPD variously as $G_3 $ or $G_2 $; these are
related to $\widetilde{E}_{2T} $ \cite{lpproc}. The connection of
these GPDs to quark orbital angular momentum was established using the
operator product expansion. Further discussion was given in \cite{yoshida}.
Below, an alternative approach based on GTMDs that connects
(\ref{e2ttildeapp}) to (\ref{lf14}) \cite{eomlir,eomlong} will be reviewed.
\end{itemize}
Both the expressions (\ref{jisr}) and (\ref{lf14}) have been employed
previously to evaluate quark orbital angular momentum in the proton
within Lattice QCD. Ji's sum rule is the traditional approach that has
been taken in numerous studies, cf., e.g.,
\cite{LHPC_1,LHPC_2,reg2004,reg2019,liu2015,etm2017,etm2020}.
On the other hand, the approach via the GTMD $F_{14} $ was used to
extend the treatment from Ji to Jaffe-Manohar quark orbital angular
momentum in \cite{jitojm,f14deriv}, with consistency of the results
for Ji quark orbital angular momentum obtained via either (\ref{jisr})
or (\ref{lf14}) demonstrated in \cite{f14deriv}. In the present
investigation, the third approach, via (\ref{e2ttildeapp}), is explored.
After elucidating the connection between (\ref{e2ttildeapp}) and
(\ref{lf14}) in the next section, the setup and realization of a
lattice calculation of the second Mellin moment of $\widetilde{E}_{2T} $
in (\ref{e2ttildeapp}) is discussed, and the result is confronted with
evaluations of Ji quark orbital angular momentum via (\ref{jisr}) and
(\ref{lf14}).

\section{Equation of motion relation connecting $\widetilde{E}_{2T} $
to orbital angular momentum}
Proton GTMDs parametrize \cite{mms} the GTMD correlator
\begin{equation}
W_{\Lambda' \Lambda }^\Gamma
= \frac{1}{2} \int
\frac{d z^- \, d^2 z_T}{(2 \pi)^3}
e^{ixP^+ z^- - i k_T\cdot z_T}
\left. \langle p', \Lambda' \mid
\bar{\psi} (-z/2) \, \Gamma \, {\cal U} \,
\psi (z/2) \color{black}
\mid p, \Lambda \rangle \right|_{z^+=0} \ \ ,
\label{gtmdcorr}
\end{equation}
where $\Lambda' , \Lambda $ denote the helicities of the proton states
carrying momenta $p' ,p$, respectively; the quark operators located at
$-z/2,z/2$ are connected by a gauge link ${\cal U}$, which for present
purposes will be taken to follow a straight-line path between the quark
operator locations. $\Gamma $ denotes a Dirac matrix structure, and the
quark longitudinal momentum fraction $x$ and transverse momentum $k_T $
are Fourier conjugate to $P^{+} z^{-} $ and $z_T $, respectively. Here,
the average of $p'$ and $p$ is denoted as $P=(p' +p)/2$; the difference,
which will be taken to be purely transverse in the following, is the
momentum transfer $\Delta_{T} = p' -p$.

The derivation of equation of motion relations between GTMD correlators
can be sketched as follows (cf.~\cite{eomlong} for details):
Consider replacing $\psi $ in (\ref{gtmdcorr}) by
$0=(iD\hspace{-0.22cm} / \ -m)\, \psi $, corresponding to the equation of
motion for the quark field. The resulting vanishing expression can be
returned to a form containing GTMD correlators by removing the derivative
from the $\psi $ field employing integration by parts. This generates two
types of terms: On the one hand, derivatives may act on the exponential
factor in (\ref{gtmdcorr}), yielding GTMD correlators multiplied by momenta;
on the other hand, derivatives may act on the gauge link ${\cal U}$,
yielding quark-gluon-quark correlators distinct from GTMD correlators.
Note that derivatives acting on the $\bar{\psi } $ field can be eliminated
by invoking also the adjoint equation of motion
$0=\bar{\psi } \, (i\overleftarrow{D}\!\!\!\! / \ +m) $; by combining
the expressions obtained using the original and adjoint quark equations
of motion symmetrically, the quark mass term can be canceled.

Carrying out these steps specifically for $\Gamma =i\sigma^{i+} \gamma^{5} $,
where $i$ is a transverse vector index, one obtains the relations
\begin{equation}
0 = ik^+\epsilon^{ij}W^{\gamma^j}_{\Lambda' \Lambda } +
\frac{\Delta^{i} }{2} W^{\gamma^+\gamma^5}_{\Lambda' \Lambda }
-i \epsilon^{ij}k^j W^{\gamma^+}_{\Lambda' \Lambda } +
{\cal M}^{i}_{\Lambda' \Lambda } \ \ ,
\label{eomgen}
\end{equation}
where it has already been used that the momentum transfer is purely
transverse for present purposes. ${\cal M}^{i}_{\Lambda' \Lambda } $
denotes a quark-gluon-quark term; $i,j$ are transverse vector indices.
These equation of motion relations among GTMD correlators imply
corresponding relations among the GTMDs that parametrize \cite{mms} them.
By judicious choice of $\Lambda' , \Lambda $ and contractions of the
transverse $i$ index, one can isolate particular GTMDs of interest.
For present purposes, the relevant combination is 
$W_{++}^\Gamma - W_{--}^\Gamma $, together with contraction with
$\Delta^{i} /\Delta_{T}^{2} $. This results in the GTMD relation
\begin{equation}
0 = -2x\left( \frac{k_T \cdot \Delta_{T} }{\Delta_{T}^{2} } F_{27}
+ F_{28} \right) +G_{14} -2 \frac{k_T^2 \Delta_{T}^{2} -
(k_T \cdot \Delta_{T} )^{2} }{M^2 \Delta_{T}^{2} } F_{14} +
\frac{\Delta^{i} }{\Delta_{T}^{2} }
\left( {\cal M}_{++}^{i} - {\cal M}_{--}^{i} \right) \ .
\label{eomgtmd}
\end{equation}
The ordering of terms in (\ref{eomgen}) and (\ref{eomgtmd}), as well as
(\ref{eomgpd}) below, is identical for easy reference. Upon integration
over $k_T $, identifying\footnote{The identification of $k_T $-integrals
of (G)TMD quantities with collinear quantities such as GPDs in general is
subject to loop corrections depending on the handling of ultraviolet
divergences on the (G)TMD vs.~the collinear side. The importance of
systematic effects arising more generally from varying treatments of
the ultraviolet divergences will be the subject of further comment below.}
the resulting GPDs \cite{mms}, one obtains the equation of motion
relation
\begin{equation}
0 = x \widetilde{E}_{2T} + \widetilde{H} -
2 \int d^2 k_T \, \frac{k_T^2 \Delta_{T}^{2} -
(k_T \cdot \Delta_{T} )^{2} }{M^2 \Delta_{T}^{2} } F_{14} +
\int d^2 k_T \frac{\Delta^{i} }{\Delta_T^2}
\left( {\cal M}^{i}_{++} - {\cal M}^{i}_{--} \right) \ .
\label{eomgpd}
\end{equation}
This relation is valid point by point in momentum fraction $x$ and
(transverse) momentum transfer $\Delta_{T} $. To arrive at quark
orbital angular momentum, one must integrate over $x$ and take the
forward limit. In that case, the quark-gluon-quark term integrates
to zero \cite{eomlong} and one finally arrives at the relation
\begin{equation}
-\int dx \int d^2 k_T \ \frac{k_T^2}{M^2 } \, F_{14} =
-\int dx\, x \widetilde{E}_{2T} - \int dx\, \widetilde{H} \ ,
\label{eomresult}
\end{equation}
demonstrating the equivalence of (\ref{lf14}) and (\ref{e2ttildeapp}).

Note that also the equivalence of these with (\ref{jisr}) can be
established within the same GTMD framework, by supplementing the
equation of motion relation discussed above with a Lorentz invariance
relation, as laid out in detail in \cite{eomlong}.

\section{Extraction of $\widetilde{E}_{2T} $ Mellin moment from
GTMD correlator}
To extract the second Mellin moment of $\widetilde{E}_{2T} $ relevant
for quark orbital angular momentum, cf.~(\ref{e2ttildeapp}), from a
GTMD correlator, one can refer to the steps that led from (\ref{eomgen})
to (\ref{eomresult}) in reverse order:
\begin{eqnarray}
L_3 + 2S_3 &=& -\int dx\, x\widetilde{E}_{2T} \\
&=& 2\int dx\, x \int d^2 k_T
\left( \frac{k_T \cdot \Delta_{T} }{\Delta_{T}^{2} } F_{27} + F_{28} \right) \\
&=& -iP^{+} \int dx\, x \int d^2 k_T \epsilon_{ij}
\frac{\Delta^{i} }{\Delta_{T}^{2} }
\left( W_{++}^{\gamma^{j} } - W_{--}^{\gamma^{j} } \right) \ ,
\label{correxp}
\end{eqnarray}
where all distributions are to be taken in the forward limit,
$\Delta_{T} \rightarrow 0$; note $k^{+} = xP^{+} $. Now, referring to the
definition of the GTMD correlator (\ref{gtmdcorr}), the factor $x$ in
(\ref{correxp}) can be replaced by a derivative with respect to the
longitudinal component of $z$, i.e.,
$x \rightarrow i\partial / \partial (z\cdot P) $ acting on the QCD
matrix element (note the additional minus sign from shifting the
derivative from the exponential factor to the matrix element via
integration by parts). Then, having done so, one can carry out the
integrations over $x$ and $k_T $ in (\ref{correxp}), which simply cancel
the integrations over $z^{-} $ and $z_T $ in the GTMD correlator
(\ref{gtmdcorr}). Formally, this results in taking the limit
$z^{-} \rightarrow 0$ and $z_T \rightarrow 0$, which must be handled with
care in view of the finite resolution implied by the lattice spacing; this
will be discussed in more detail below. Furthermore, in the forward limit,
$2(\Delta^{i} / \Delta_{T}^{2} ) f^i = (\partial / \partial \Delta^{i} ) f^i $
for any vector function $f$ which vanishes at least linearly in that limit.
Thus, one arrives at
\begin{eqnarray}
& & \frac{L_3 +2S_3 }{n} = \label{e2tme} \\
& & \left. \frac{1}{n} \epsilon_{ij}
\ \frac{1}{2} \ \frac{\partial }{\partial (z\cdot P)} \
\frac{\partial }{\partial \Delta^{i} } \left\langle P+\Delta_{T} /2 , +
\right| \bar{\psi } (-z/2) \gamma^{j} \, {\cal U} \, \psi (z/2)
\left| P-\Delta_{T} /2 , + \right\rangle
\right|_{z^{+} =0, \Delta_{T} =0, z^{-} \rightarrow 0 ,z_T \rightarrow 0}
\ ,
\nonumber
\end{eqnarray}
where it has also been used that the 
$W_{++}^{\gamma^{j} } $ and $W_{--}^{\gamma^{j} } $ contributions
are identical (up to their sign), and therefore it is sufficient to
calculate the former and multiply by a factor 2. The normalization by
the number of valence quarks $n$ serves to cancel the renormalization
factor of the quark bilinear operator; it can be obtained from the
matrix element
\begin{equation}
n = \left. \frac{1}{2 P^j } \left\langle P^j , + \right|
\bar{\psi } (-z/2) \gamma^{j} \, {\cal U} \, \psi (z/2)
\left| P^j , + \right\rangle
\right|_{z^{+} =0, z^{-} \rightarrow 0 ,z_T \rightarrow 0 }
\label{e2tnorm}
\end{equation}
(no summation over $j$ implied); in this matrix element, the proton
momentum is in the $j$-direction, orthogonal to $P$ defining the
longitudinal direction in (\ref{e2tme}). Alternatively, one can
invoke invariance of (\ref{e2tnorm}) under rotation of the $j$-direction
to align with the direction of $P$ in (\ref{e2tme}). This was done in
practice, in order to avoid having to perform calculations for additional
proton momenta; recording data for an additional Dirac matrix structure,
$\gamma^{\text{longitudinal} } $, is, in comparison, much less expensive.

\section{Setup of lattice calculation and result}
In the Lorentz frame in which the GTMD correlator (\ref{gtmdcorr}) is
originally defined, the quark bilocal operator contains temporal separations.
This frame is therefore not suitable for a lattice calculation. The lattice
calculation instead is performed in a boosted frame in which the separation
in the operator is purely spatial. To facilitate the connection between the
frames, it is useful to formulate the problem in terms of the invariants
$z\cdot P$ and $z^2 $:
\begin{eqnarray}
\begin{array}{llll}
\mbox{Original frame:} & z^{+} =0 , & z\cdot P = z^{-} P^{+} , & z^2 = -z_T^2 \\
\mbox{Lattice frame:} & z_{0} =0 , & z\cdot P = -z_{3} P_{3} , &
z^2 = -z_3^2 -z_T^2 \ \ ,
\end{array}
\end{eqnarray}
where, as before, the spatial direction of the proton momentum $P$ has been
taken to define the $3$-direction. Also, in the following, the momentum
transfer $\Delta_{T} $ will point in the $1$-direction, and the transverse
component of the operator separation $z_T $ will point in the $2$-direction.

\begin{figure}
\begin{center}
\includegraphics[width=7cm]{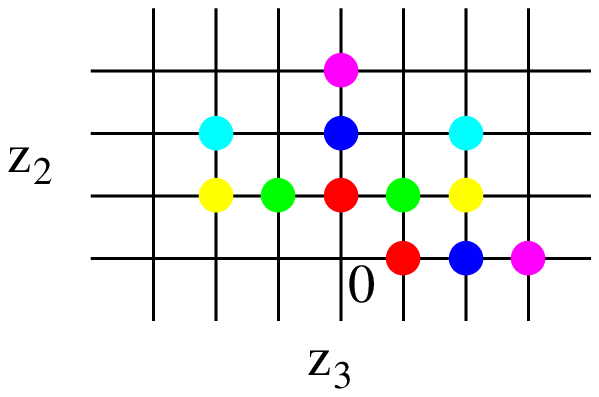}
\end{center}
\vspace{-0.5cm}
\caption{Pairs of values of $z_3 $ at constant $z^2 =-z_3^2 -z_2^2 $ used
to evaluate the finite difference approximating the derivative with respect
to $z\cdot P$ in (\ref{e2tme}). Each pair is distinguished by its color.
The spacing of the grid is the lattice spacing $a$ (note that the displayed
grid is not the underlying lattice itself, but it is a grid of relative
distances on that lattice). The final result for $(L_3 +2S_3 )/n$ is
determined by the $z^2 = -a^2 $ case (red dots).}
\label{figdiff}
\end{figure}

Obtaining the form (\ref{e2tme}) from (\ref{correxp}), one integrates
over the quark momenta $k_T $ and $k^{-} $. Formally, this implies
the limits $z^{-} \rightarrow 0 ,z_T \rightarrow 0$ in (\ref{e2tme}),
if one were to integrate over all momenta without any cutoff. However,
the finite resolution implied by the lattice spacing $a$ implies a cutoff
on momenta, or, formulated in $z$-space, quark operator separations
smaller than $|z|=a$ cannot be resolved. Thus, the aforementioned limits
$z^{-} \rightarrow 0 ,z_T \rightarrow 0$ will be understood to mean
evaluation at fixed $-z^2 =a^2 $; likewise, the derivative with respect
to $z_3 $ (in the lattice frame) in (\ref{e2tme}) will be taken to mean a
finite difference over the minimally resolved distance $a$. In the numerical
calculations to be presented below, a range of $|z|$ was studied, with the
case $|z|=a$ determining the final results. Fig.~\ref{figdiff} shows pairs
of values of $z_3 $ at constant $z^2 $ used to evaluate the finite
difference approximating the derivative with respect to $z\cdot P$ in
(\ref{e2tme}). Consistently, also the denominator $n$ in (\ref{e2tme}) is
evaluated at the same $z^2 $, matching the operators in numerator and
denominator at finite lattice spacing.

Note that this (gauge-invariant) momentum cutoff scheme, with $z^2 $
effectively defining the cutoff on momentum integrations, differs from the
perturbative $\overline{\mbox{MS} } $ renormalization scheme. Conversion of
the results to the latter would require a matching factor that has
currently not yet been determined. Systematic deviations between the
two schemes have, however, been estimated to be minor \cite{sun} when the
momentum cutoff is comparable to the renormalization scale. The
numerical results presented below corroborate that the systematic
uncertainty associated with the connection to the $\overline{\mbox{MS} } $
scheme is not dominant compared to other uncertainties of the calculation.

In addition to the derivative with respect to $z\cdot P$, (\ref{e2tme})
calls for a derivative of the matrix element with respect to (transverse)
momentum transfer. This derivative was realized using a direct derivative
method \cite{rome,nhasan,f14deriv} in order to avoid any systematic bias
in its evaluation. The treatment is identical to that carried out in
\cite{f14deriv}, and the reader is referred to the detailed description
there for specifics.

\begin{figure}
\begin{center}
\includegraphics[width=7.3cm]{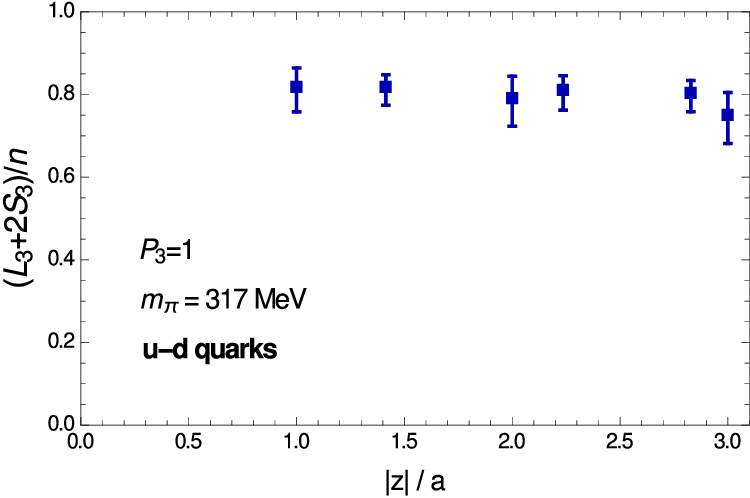}
\hspace{0.3cm}
\includegraphics[width=7.3cm]{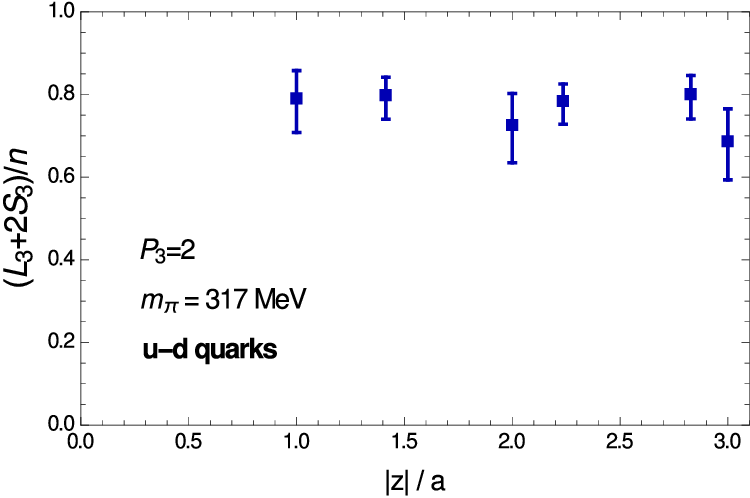}
\end{center}
\caption{The combination $L_3 +2S_3 $, in units of the number of
valence quarks $n$, obtained through the second Mellin moment of the
twist-3 GPD $\widetilde{E}_{2T} $, as a function of the spatial
cutoff $|z|$ used in evaluating (\ref{e2tme}), cf.~discussion in
main text. The two panels display two different proton momenta
$P_3 $ (values quoted in panel legends are given in units of
$2\pi /\ell $, cf.~main text).}
\label{figzdep}
\end{figure}

Numerical data for the ratio (\ref{e2tme}) were obtained utilizing a
2+1-flavor clover fermion ensemble on a $32^3 \times 96$ lattice with
spacing $a=0.114\, \mbox{fm} = (1.73 \, \mbox{GeV} )^{-1} $. The pion mass
on this ensemble is $m_{\pi } =317\, \mbox{MeV} $, and the source-sink
separation employed to construct three-point functions was
$10\, a = 1.14\, \mbox{fm} $. Two proton momenta were explored,
$P_3 = 2\pi /\ell ,\, 4\pi /\ell $, where $\ell =32\, a$ denotes the spatial
lattice extent. Note that non-zero proton momentum is required to evaluate
(\ref{e2tnorm}) (or spatial rotations thereof). The Ji definition of
longitudinal orbital angular momentum (and likewise longitudinal spin) is
boost-invariant, but lattice calculations at different $P_3 $ may deviate
from one another, e.g., as a result of discretization artefacts that will
tend to increase with rising $P_3 $. Fig.~\ref{figzdep} displays results
for $(L_3 + 2S_3 )/n$ for the two proton momenta, as a function of the
spatial cutoff $|z|$ used in evaluating (\ref{e2tme}). The results are
remarkably stable with respect to $|z|$ and agree within uncertainties for
the two proton momenta, as expected. Of course, $L_3 $ is expected to evolve
nontrivially with the ultraviolet scale. However, at the given lattice
spacing, there appears to be little ambiguity in the extraction of
$(L_3 + 2S_3 )/n$, in view of its stability with respect to the
implementation of the momentum cutoff.

To extract the final estimate for Ji longitudinal quark orbital angular
momentum in the proton, the spin contribution $2S_3 $ must be subtracted
from (\ref{e2tme}). This contribution was previously evaluated on the same
lattice ensemble in \cite{axial}. To achieve a combination of data that is
as consistent as possible, the results from \cite{axial} were evaluated
at matching fixed source-sink separation $10\, a$ (rather than using the
extrapolation to the ground state also available there), yielding
$2S_3 = 1.18(2)$. It should be noted that, whereas this result does
not depend on renormalization scheme or scale (in the isovector case
considered here), the handling of discretization effects in arriving
at it does not fully coincide with the one implied by the momentum cutoff
scheme adopted above in extracting $(L_3 + 2S_3 )/n$. However, being
performed at a single lattice spacing, no quantitative assessment of the
discretization error of the present calculation is possible in any case.
In addition, the $|z|=a$ data from Fig.~\ref{figzdep} were linearly
extrapolated to $P_3 =0$ for the purpose of combining the data. The
final estimate of the isovector longitudinal Ji quark orbital angular
momentum in the proton obtained in this manner is
\begin{equation}
L_3 /n = -0.34(6) \ .
\label{finresult}
\end{equation}

\begin{figure}
\begin{center}
\includegraphics[width=5cm]{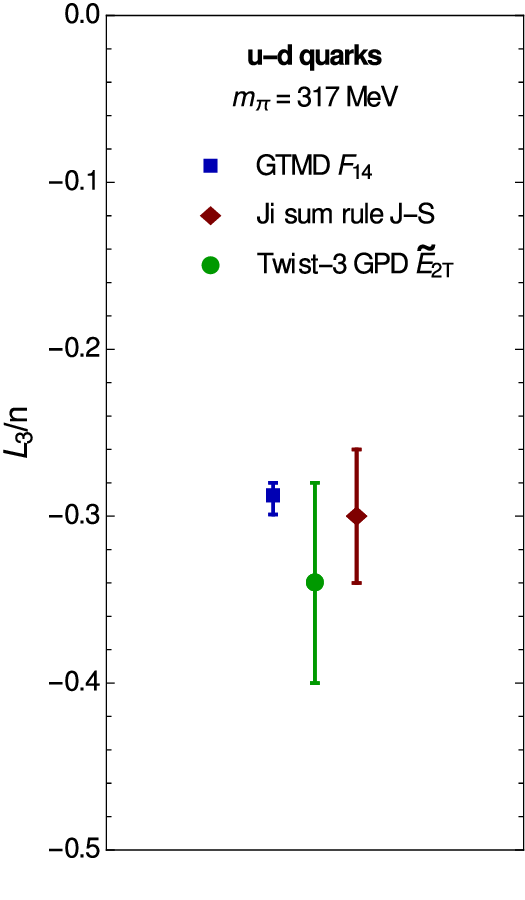}
\end{center}
\caption{Isovector longitudinal Ji quark orbital angular momentum in the
proton, in units of the number of valence quarks $n$, obtained in lattice
calculations utilizing the three approaches given by eqs.~(\ref{jisr}),
(\ref{lf14}) and (\ref{e2ttildeapp}) discussed in the introduction.}
\label{figfinal}
\end{figure}

\section{Discussion}
Fig.~\ref{figfinal} places the result (\ref{finresult}) into the context
of alternative evaluations of isovector longitudinal Ji quark orbital
angular momentum in the proton based on both the GTMD $F_{14} $,
cf.~(\ref{lf14}), as well as Ji's sum rule, cf.~(\ref{jisr}). The
result employing the GTMD $F_{14} $ was obtained \cite{f14deriv} on the
same lattice ensemble as utilized in the present investigation, in the
same quark momentum cutoff scheme as used in evaluating (\ref{e2tme}).
The result employing Ji's sum rule was obtained by interpolating data
from \cite{LHPC_2} to the same pion mass as employed here; these data
are given in the $\overline{\mbox{MS} } $ scheme at scale
$\mu = 2\, \mbox{GeV} $.

The various determinations agree within the statistical uncertainties;
the different systematics inherent in the distinct approaches, including
the schemes in which the ultraviolet divergences are handled, appear to
only influence the results for the quark orbital angular momentum to an
insignificant extent, compared to the statistical fluctuations. The
magnitude of those fluctuations in the determination of quark orbital
angular momentum via the twist-3 GPD $\widetilde{E}_{2T} $ is notable. It
is much larger than the one using the GTMD $F_{14} $, obtained on the
same ensemble with the same statistics (the Ji sum rule result cannot
be directly compared in the same way concerning the statistical fluctuations,
since it was obtained on different ensembles employing significantly
smaller sets of samples). The main reason for the comparatively large
statistical uncertainty of the result obtained using the twist-3 GPD
$\widetilde{E}_{2T} $ is that it is realized through a difference of
large numbers, $L_3 = (L_3 + 2S_3 ) - 2S_3 $, canceling a large part
of the signal and thus enhancing the relative uncertainty. Also a
less than optimal cancellation of fluctuations in the numerator and
denominator of (\ref{e2tme}) may play a role, when the matrix element
in the denominator (\ref{e2tnorm}) is rotated to align the $j$-direction
with the direction of $P$ in (\ref{e2tme}), as remarked after
eq.~(\ref{e2tnorm}).

Nonetheless, the present study demonstrates the feasibility of extracting
quark orbital angular momentum in the proton from the twist-3 GPD
$\widetilde{E}_{2T} $, albeit requiring higher computational effort compared
to other approaches to achieve comparable accuracy. Presumably also
phenomenological studies employing the GPD $\widetilde{E}_{2T} $ would
have to grapple with the numerical challenges inherent in the difference
of large numbers required to determine the quark orbital angular momentum
in the proton.

\section*{Acknowledgments}
Computations were performed using resources provided by the U.S.~DOE
Office of Science through the National Energy Research Scientific
Computing Center (NERSC), a DOE Office of Science User Facility
located at Lawrence Berkeley National Laboratory, under Contract
No.~DE-AC02-05CH11231, as well as through facilities of the USQCD
Collaboration, employing the Chroma and Qlua software suites.
R.~Edwards, B.~Jo\'{o} and K.~Orginos are acknowledged for providing
the clover ensemble analyzed in this work, which was generated using
resources provided by XSEDE (supported by National Science Foundation
Grant No.~ACI-1053575). M.E., S.L., J.N., and A.P.~are supported by the
U.S.~DOE Office of Science, Office of Nuclear Physics, through grants
DE-FG02-96ER40965, DE-SC0016286, DE-SC-0011090 and DE-SC0023116,
respectively. M.R.~was supported under the RWTH Exploratory Research
Space (ERS) grant PF-JARA-SDS005 and MKW NRW under the funding code
NW21-024-A. S.M.~is supported by the U.S.~DOE Office of Science,
Office of High Energy Physics, under Award Number DE-SC0009913.
S.S.~is supported by the National Science Foundation under CAREER
Award PHY-1847893. This work was furthermore supported by the U.S.~DOE
through the TMD Topical Collaboration.


\begin{thebibliography}{99}
\bibitem{emc1} J.~Ashman {\it et al.} [European Muon Collaboration],
Phys. Lett. {\bf B206} (1988) 364.
\bibitem{emc2} J.~Ashman {\it et al.} [European Muon Collaboration],
Nucl. Phys. {\bf B328} (1989) 1.
\bibitem{jidecomp} X.~Ji, Phys. Rev. Lett. {\bf 78} (1997) 610.
\bibitem{jmdecomp} R.~Jaffe and A.~Manohar,
Nucl. Phys. {\bf B337} (1990) 509.
\bibitem{hatta} Y.~Hatta, Phys. Lett. {\bf B708} (2012) 186.
\bibitem{lorce} C.~Lorc\'e and B.~Pasquini,
Phys. Rev. {\bf D 84} (2011) 014015.
\bibitem{mms} S.~Mei\ss ner, A.~Metz and M.~Schlegel,
JHEP {\bf 0908} (2009) 056.
\bibitem{burk} M.~Burkardt, Phys. Rev. {\bf D 88}, 014014 (2013).
\bibitem{jist} X.~Ji, X.~Xiong and F.~Yuan,
Phys. Rev. Lett. {\bf 109}, 152005 (2012).
\bibitem{eomlir} A.~Rajan, A.~Courtoy, M.~Engelhardt and S.~Liuti,
Phys. Rev. {\bf D 94}, 034041 (2016).
\bibitem{lpproc} C.~Lorc\'e and B.~Pasquini,
Int. J. Mod. Phys. Conf. Ser. {\bf 20} (2012) 84.
\bibitem{eomlong} A.~Rajan, M.~Engelhardt and S.~Liuti,
Phys. Rev. {\bf D 98}, 074022 (2018).
\bibitem{polyakov1} M.~Penttinen, M.~Polyakov, A.~Shuvaev and M.~Strikman,
Phys. Lett. {\bf B491} (2000) 96.
\bibitem{polyakov2} D.~Kiptily and M.~Polyakov,
Eur. Phys. J. {\bf C 37} (2004) 105.
\bibitem{yoshida} Y.~Hatta and S.~Yoshida, JHEP {\bf 10} (2012) 080.
\bibitem{LHPC_1} P.~H\"agler {\it et al.} [LHP Collaboration],
Phys. Rev. {\bf D 77}, 094502 (2008).
\bibitem{LHPC_2} J.~D.~Bratt {\it et al.} [LHP Collaboration],
Phys. Rev. {\bf D 82}, 094502 (2010).
\bibitem{reg2004} M.~G\"ockeler, R.~Horsley, D.~Pleiter, P.~E.~L.~Rakow,
A.~Sch\"afer, G.~Schierholz and W.~Schroers [QCDSF Collaboration],
Phys. Rev. Lett. {\bf 92}, 042002 (2004).
\bibitem{reg2019} G.~Bali, S.~Collins, M.~G\"ockeler, R.~R\"odl,
A.~Sch\"afer and A.~Sternbeck [RQCD Collaboration],
Phys. Rev. {\bf D 100}, 014507 (2019).
\bibitem{liu2015} M.~Deka, T.~Doi, Y.-B.~Yang, B.~Chakraborty, S.~J.~Dong,
T.~Draper, M.~Glatzmaier, M.~Gong, H.-W.~Lin, K.-F.~Liu, D.~Mankame,
N.~Mathur and T.~Streuer, Phys. Rev. {\bf D 91} (2015), 014505 (2015).
\bibitem{etm2017} C.~Alexandrou, M.~Constantinou, K.~Hadjiyiannakou,
K.~Jansen, C.~Kallidonis, G.~Koutsou, A.~Vaquero Avil\'es-Casco and C.~Wiese,
Phys. Rev. Lett. {\bf 119}, 142002 (2017).
\bibitem{etm2020} C.~Alexandrou, S.~Bacchio, M.~Constantinou, J.~Finkenrath,
K.~Hadjiyiannakou, K.~Jansen, G.~Koutsou, H.~Panagopoulos and G.~Spanoudes,
Phys. Rev. {\bf D 101}, 094513 (2020).
\bibitem{jitojm} M.~Engelhardt, Phys. Rev. {\bf D 95}, 094505 (2017).
\bibitem{sun} M.~Ebert, J.~Michel, I.~Stewart and Z.~Sun,
JHEP {\bf 07} (2022) 129.
\bibitem{rome} G.~M.~de~Divitiis, R.~Petronzio and N.~Tantalo,
Phys. Lett. {\bf B718}, 589 (2012).
\bibitem{nhasan} N.~Hasan, J.~R.~Green, S.~Meinel, M.~Engelhardt, S.~Krieg,
J.~Negele, A.~Pochinsky and S.~Syritsyn,
Phys. Rev. {\bf D 97}, 034504 (2018).
\bibitem{f14deriv} M.~Engelhardt, J.~R.~Green, N.~Hasan, S.~Krieg, S.~Meinel,
J.~Negele, A.~Pochinsky and S.~Syritsyn,
Phys. Rev. {\bf D 102} (2020) 074505.
\bibitem{axial} J.~R.~Green, N.~Hasan, S.~Meinel, M.~Engelhardt, S.~Krieg,
J.~Laeuchli, J.~Negele, K.~Orginos, A.~Pochinsky and S.~Syritsyn,
Phys. Rev. {\bf D 95} (2017) 114502.
\end{thebibliography}
\end{document}